\documentclass[aps,prX,twocolumn, superscriptaddress,floatfix,showpacs]{revtex4-1}

\usepackage{notes2bib}
\usepackage{graphicx}
\usepackage{gensymb}
\usepackage{url}
\usepackage{graphicx}
\usepackage{amsmath}
\usepackage{latexsym}
\usepackage{amssymb}
\usepackage{color}
\usepackage[normalem]{ulem}
\usepackage[separate-uncertainty]{siunitx}
\DeclareSIUnit[number-unit-product = {\,}]\clight{$c$}
\usepackage{microtype}
\usepackage{dcolumn}

\usepackage{diagbox}

\newcommand{\bToD}{$B^+ \rightarrow D^-\pi^+ \ell^+\nu$~}
\newcommand{\bZToD}{$B^0 \rightarrow \bar{D}^0\pi^- \ell^+\nu$~}
\newcommand{\bToDStar}{$B^+ \rightarrow D^{*-}\pi^+ \ell^+\nu$~}
\newcommand{\bZToDStar}{$B^0 \rightarrow \bar{D}^{*0}\pi^- \ell^+\nu$~}

\newcommand{\bToDNS}{$B^+ \rightarrow D^-\pi^+ \ell^+\nu$}
\newcommand{\bZToDNS}{$B^0 \rightarrow \bar{D}^0\pi^- \ell^+\nu$}
\newcommand{\bToDStarNS}{$B^+ \rightarrow D^{*-}\pi^+ \ell^+\nu$}
\newcommand{\bZToDStarNS}{$B^0 \rightarrow \bar{D}^{*0}\pi^- \ell^+\nu$}

\newcommand{\ra}[1]{\renewcommand{\arraystretch}{#1}}

\begin{document}


\preprint{\vbox{ \hbox{   }
						\hbox{Belle DRAFT {\it YY-NN}}
                        \hbox{Intended for {\it Phys.Rev.D}}
                        \hbox{Author: A. Vossen}
                        \hbox{Committee: Y. Sakai(chair),}
                        \hbox{T. Nanut, O. Hartbrich, }
}}
\title{\quad\\[0.5cm]Measurement of the branching fraction of $B \rightarrow D^{(*)}\pi \ell\nu$ at Belle using hadronic tagging in fully reconstructed events}


\date{\today}
\noaffiliation
\affiliation{University of the Basque Country UPV/EHU, 48080 Bilbao}
\affiliation{Beihang University, Beijing 100191}
\affiliation{Brookhaven National Laboratory, Upton, New York 11973}
\affiliation{Budker Institute of Nuclear Physics SB RAS, Novosibirsk 630090}
\affiliation{Faculty of Mathematics and Physics, Charles University, 121 16 Prague}
\affiliation{University of Cincinnati, Cincinnati, Ohio 45221}
\affiliation{Deutsches Elektronen--Synchrotron, 22607 Hamburg}
\affiliation{Duke University, Durham, North Carolina 27708}
\affiliation{University of Florida, Gainesville, Florida 32611}
\affiliation{Justus-Liebig-Universit\"at Gie\ss{}en, 35392 Gie\ss{}en}
\affiliation{Gifu University, Gifu 501-1193}
\affiliation{II. Physikalisches Institut, Georg-August-Universit\"at G\"ottingen, 37073 G\"ottingen}
\affiliation{SOKENDAI (The Graduate University for Advanced Studies), Hayama 240-0193}
\affiliation{Gyeongsang National University, Chinju 660-701}
\affiliation{Hanyang University, Seoul 133-791}
\affiliation{University of Hawaii, Honolulu, Hawaii 96822}
\affiliation{High Energy Accelerator Research Organization (KEK), Tsukuba 305-0801}
\affiliation{J-PARC Branch, KEK Theory Center, High Energy Accelerator Research Organization (KEK), Tsukuba 305-0801}
\affiliation{IKERBASQUE, Basque Foundation for Science, 48013 Bilbao}
\affiliation{Indian Institute of Technology Bhubaneswar, Satya Nagar 751007}
\affiliation{Indian Institute of Technology Guwahati, Assam 781039}
\affiliation{Indian Institute of Technology Hyderabad, Telangana 502285}
\affiliation{Indian Institute of Technology Madras, Chennai 600036}
\affiliation{Indiana University, Bloomington, Indiana 47408}
\affiliation{Institute of High Energy Physics, Chinese Academy of Sciences, Beijing 100049}
\affiliation{Institute of High Energy Physics, Vienna 1050}
\affiliation{INFN - Sezione di Napoli, 80126 Napoli}
\affiliation{INFN - Sezione di Torino, 10125 Torino}
\affiliation{Advanced Science Research Center, Japan Atomic Energy Agency, Naka 319-1195}
\affiliation{J. Stefan Institute, 1000 Ljubljana}
\affiliation{Institut f\"ur Experimentelle Teilchenphysik, Karlsruher Institut f\"ur Technologie, 76131 Karlsruhe}
\affiliation{Kennesaw State University, Kennesaw, Georgia 30144}
\affiliation{King Abdulaziz City for Science and Technology, Riyadh 11442}
\affiliation{Department of Physics, Faculty of Science, King Abdulaziz University, Jeddah 21589}
\affiliation{Korea Institute of Science and Technology Information, Daejeon 305-806}
\affiliation{Korea University, Seoul 136-713}
\affiliation{Kyoto University, Kyoto 606-8502}
\affiliation{Kyungpook National University, Daegu 702-701}
\affiliation{LAL, Univ. Paris-Sud, CNRS/IN2P3, Universit\'{e} Paris-Saclay, Orsay}
\affiliation{\'Ecole Polytechnique F\'ed\'erale de Lausanne (EPFL), Lausanne 1015}
\affiliation{P.N. Lebedev Physical Institute of the Russian Academy of Sciences, Moscow 119991}
\affiliation{Faculty of Mathematics and Physics, University of Ljubljana, 1000 Ljubljana}
\affiliation{Ludwig Maximilians University, 80539 Munich}
\affiliation{Luther College, Decorah, Iowa 52101}
\affiliation{University of Malaya, 50603 Kuala Lumpur}
\affiliation{University of Maribor, 2000 Maribor}
\affiliation{Max-Planck-Institut f\"ur Physik, 80805 M\"unchen}
\affiliation{School of Physics, University of Melbourne, Victoria 3010}
\affiliation{University of Mississippi, University, Mississippi 38677}
\affiliation{Moscow Physical Engineering Institute, Moscow 115409}
\affiliation{Moscow Institute of Physics and Technology, Moscow Region 141700}
\affiliation{Graduate School of Science, Nagoya University, Nagoya 464-8602}
\affiliation{Universit\`{a} di Napoli Federico II, 80055 Napoli}
\affiliation{Nara Women's University, Nara 630-8506}
\affiliation{National Central University, Chung-li 32054}
\affiliation{National United University, Miao Li 36003}
\affiliation{Department of Physics, National Taiwan University, Taipei 10617}
\affiliation{H. Niewodniczanski Institute of Nuclear Physics, Krakow 31-342}
\affiliation{Nippon Dental University, Niigata 951-8580}
\affiliation{Niigata University, Niigata 950-2181}
\affiliation{Novosibirsk State University, Novosibirsk 630090}
\affiliation{Osaka City University, Osaka 558-8585}
\affiliation{Pacific Northwest National Laboratory, Richland, Washington 99352}
\affiliation{Panjab University, Chandigarh 160014}
\affiliation{University of Pittsburgh, Pittsburgh, Pennsylvania 15260}
\affiliation{Theoretical Research Division, Nishina Center, RIKEN, Saitama 351-0198}
\affiliation{University of Science and Technology of China, Hefei 230026}
\affiliation{Showa Pharmaceutical University, Tokyo 194-8543}
\affiliation{Soongsil University, Seoul 156-743}
\affiliation{Stefan Meyer Institute for Subatomic Physics, Vienna 1090}
\affiliation{Sungkyunkwan University, Suwon 440-746}
\affiliation{School of Physics, University of Sydney, New South Wales 2006}
\affiliation{Department of Physics, Faculty of Science, University of Tabuk, Tabuk 71451}
\affiliation{Tata Institute of Fundamental Research, Mumbai 400005}
\affiliation{Excellence Cluster Universe, Technische Universit\"at M\"unchen, 85748 Garching}
\affiliation{Department of Physics, Technische Universit\"at M\"unchen, 85748 Garching}
\affiliation{Department of Physics, Tohoku University, Sendai 980-8578}
\affiliation{Earthquake Research Institute, University of Tokyo, Tokyo 113-0032}
\affiliation{Department of Physics, University of Tokyo, Tokyo 113-0033}
\affiliation{Tokyo Institute of Technology, Tokyo 152-8550}
\affiliation{Tokyo Metropolitan University, Tokyo 192-0397}
\affiliation{Virginia Polytechnic Institute and State University, Blacksburg, Virginia 24061}
\affiliation{Wayne State University, Detroit, Michigan 48202}
\affiliation{Yamagata University, Yamagata 990-8560}
\affiliation{Yonsei University, Seoul 120-749}
\author{A.~Vossen}\affiliation{Duke University, Durham, North Carolina 27708}\affiliation{Indiana University, Bloomington, Indiana 47408} 
  \author{I.~Adachi}\affiliation{High Energy Accelerator Research Organization (KEK), Tsukuba 305-0801}\affiliation{SOKENDAI (The Graduate University for Advanced Studies), Hayama 240-0193} 
  \author{K.~Adamczyk}\affiliation{H. Niewodniczanski Institute of Nuclear Physics, Krakow 31-342} 
  \author{H.~Aihara}\affiliation{Department of Physics, University of Tokyo, Tokyo 113-0033} 
  \author{S.~Al~Said}\affiliation{Department of Physics, Faculty of Science, University of Tabuk, Tabuk 71451}\affiliation{Department of Physics, Faculty of Science, King Abdulaziz University, Jeddah 21589} 
\author{D.~M.~Asner}\affiliation{Brookhaven National Laboratory, Upton, New York 11973} 
  \author{V.~Aulchenko}\affiliation{Budker Institute of Nuclear Physics SB RAS, Novosibirsk 630090}\affiliation{Novosibirsk State University, Novosibirsk 630090} 
  \author{T.~Aushev}\affiliation{Moscow Institute of Physics and Technology, Moscow Region 141700} 
  \author{R.~Ayad}\affiliation{Department of Physics, Faculty of Science, University of Tabuk, Tabuk 71451} 
  \author{I.~Badhrees}\affiliation{Department of Physics, Faculty of Science, University of Tabuk, Tabuk 71451}\affiliation{King Abdulaziz City for Science and Technology, Riyadh 11442} 
% ...
\collaboration{The Belle Collaboration}

\noaffiliation
\begin{abstract}
We report a measurement of the branching fractions  of the decays $B \rightarrow D^{(*)}\pi \ell\nu$. The analysis uses  772$\times 10^6$ $B\bar{B}$ pairs produced in  $e^+e^-\rightarrow \Upsilon(4S)$ data recorded by the Belle experiment at the KEKB asymmetric-energy $e^+e^-$ collider. The tagging $B$ meson in the decay is fully reconstructed in a hadronic decay mode. On the signal side, we reconstruct the decay $B \rightarrow D^{(*)}\pi \ell\nu$ $(\ell=e,\mu)$. The measured branching fractions are 
 $\mathcal{B}($\bToD$)$ = [4.55 $\pm$ 0.27 (stat.) $\pm$ 0.39 (syst.)]$\times 10^{-3}$, 
  $\mathcal{B}($\bZToD$)$ = [4.05 $\pm$ 0.36 (stat.) $\pm$   0.41 (syst.)]$\times 10^{-3}$, 
$\mathcal{B}($\bToDStar$)$ = [6.03 $\pm$ 0.43 (stat.) $\pm$  0.38 (syst.)]$\times 10^{-3}$, and
 $\mathcal{B}($\bZToDStar$)$ = [6.46 $\pm$ 0.53 (stat.) $\pm$  0.52 (syst.)]$\times 10^{-3}$.
 These are in good agreement with the current world average values.
\end{abstract}


\maketitle

\tighten

{\renewcommand{\thefootnote}{\fnsymbol{footnote}}}
\setcounter{footnote}{0}

\section{Introduction}
Semileptonic decays of $B$ mesons are an important tool for precision measurements of CKM matrix elements and precision tests of the electroweak sector of the standard model. An important recent development was the observation of a more than 3$\sigma$ deviation between the standard model expectation for $R(D^{(*)})$~\cite{Fajfer:2012vx,Amhis:2016xyh} and the combined experimental results from Babar~\cite{Lees:2013uzd,Lees:2012xj},  
Belle~\cite{Huschle:2015rga,Sato:2016svk,Hirose:2016wfn} and LHCb \cite{Aaij:2015yra,Aaij:2017deq}.
Here, $R(D^{(*)})$ is defined as the ratio of the branching fraction ($\mathcal{B}$)  of $B\rightarrow D^{(*)}\tau \nu$ and $B\rightarrow D^{(*)}l \nu,$ $(\ell=e, \mu)$.
We report on a new measurement of $B \rightarrow D^{(*)}\pi \ell\nu$, which is important as a background for $B \rightarrow D^{(*)}  \tau \nu $ decays, and in its own right, as a vehicle to
 understand high-multiplicity semi-leptonic $B$ decays. 
The process $B \rightarrow D^{(*)}\pi \ell\nu$ proceeds predominantly via $B\rightarrow (D^{**}\rightarrow D^{(*)} \pi ) \ell\nu$, where $D^{**}$ is an orbitally excited $(L=1)$ charmed meson. 
The $D^{**}$ mass-spectrum contains two doublets of states having light-quark total angular momentum $j_q=\frac{1}{2}$ and $j_q=\frac{3}{2}$~\cite{Isgur:1991wq}. All states can decay  via $D^{**}\rightarrow D^*\pi$, while the $2^+$ state can also decay via $D^{**}\rightarrow D \pi$.
Since the $D^{**}$ masses are not far from threshold, and the $j_q=\frac{3}{2}$ have a significant D-wave component, these states are narrow and were observed with a typical width of about $20$\,MeV~\cite{Liventsev:2007rb,Aubert:2008zc,Aubert:2008ea}. On the other hand, the states with $j_q=\frac{1}{2}$ decay mainly via S-wave and are therefore expected to be broad resonances with a width of several hundred MeV~\cite{Isgur:1991wq,Leibovich:1997em}. Compared to previous measurements of $\mathcal{B}(B\rightarrow D^{(*)}\pi \ell \nu)$ at Belle~\cite{Liventsev:2007rb}, the analysis presented in this report benefits from the use of the full  Belle dataset, containing  772$\times 10^6$ $B\bar{B}$ pairs, recorded at the $\Upsilon(4S)$ resonance, an improved hadronic-tagging method, and a direct extraction of the branching fractions using a fit to Monte-Carlo templates.


\section{Experimental Apparatus}
The Belle experiment~\cite{BelleDetector} at the KEKB storage ring~\cite{KEKB} recorded about 1\,ab$^{-1}$ of $e^+e^-$ annihilation data. The data were taken mainly at the $\Upsilon(4S)$ resonance at $\sqrt{s}=10.58$\,GeV, but also at $\Upsilon(1S)$ to $\Upsilon(5S)$ resonances and at $\sqrt{s}=10.52$\,GeV.
The Belle instrumentation used in this analysis includes the central drift chamber (CDC) and the silicon vertex detector, which provides precision tracking for tracks in the polar-angle range $17.0\,\degree < \theta_\textrm{lab}< 150.0\,\degree$, 
and the electromagnetic calorimeter (ECL) covering the same range. The polar angle $\theta_\textrm{lab}$ is measured with respect to the $z$-axis, which is anti-parallel to the $e^+$ beam.
Charged particle identification is performed using specific ionization measurements in the CDC, time-of-flight information from the interaction point (IP) to a barrel of scintillators, light yield in an array of aerogel Cherenkov counters in the barrel and the forward endcap, as well as a muon- and $K_L^0$-identification system in the return yoke of the superconducting solenoid, which provides a 1.5\,T magnetic field.

\section{Analysis}
The analysis strategy is based on fully reconstructing one tagging $B$ meson in a hadronic mode, then, using the rest of the event, reconstructing the signal mode with the exception of the $\nu$, which escapes undetected. Since the rest of the event has been reconstructed, it is possible to infer the escaped neutrino invariant mass $M_\nu$ from the kinematic constraints of the initial $e^+e^-$ collision. The distribution of $M^2_\nu$ is then fitted with Belle Monte Carlo (MC) simulation templates to derive the branching fraction of interest. Simulations  in this analysis use Pythia~\cite{Sjostrand:2006za} and EvtGen~\cite{Lange:2001uf} for the event generation, and GEANT3~\cite{Brun:1987ma} for the detector response. The simulation treats all $B \rightarrow D^{(*)}\pi \ell \nu$ decays as proceeding through a $B\rightarrow D^{**}$ decay, which is simulated using
the model of Leibovich-Ligeti-Stewart-Wise (LLSW)~\cite{Leibovich:1997em}. By comparing known processes, we correct the simulation of the detector for the efficiency of the particle identification of charged tracks,  $\pi^0$ and $K^0_S$ mesons as well as the misidentification probabilities of charged tracks. These corrections are dependent on the kinematics of the respective particles.
We reweight the simulation of underlying physical processes to account for newly measured values of branching fractions and related parameters. In particular, we use the latest world-average values of $D$ and $B$ meson branching fractions~\cite{Olive:2016xmw} as well as $D^*$~\cite{Amhis:2016xyh} and $D^{**}$ form factors~\cite{Leibovich:1997em}.

\subsection{$B_\textrm{tag}$ selection}
A neural-network-based multivariate classifier, as implemented in the  \texttt{NeuroBayes} package~\cite{Feindt:2006pm,Feindt:2011mr}, is used to fully reconstruct $B$ mesons that decay hadronically. The algorithm considers 17 final states for charged $B$ candidates and 15 final states for neutral $B$ candidates. Incorporating the subsequent hadronic decays and $J/\psi$ leptonic decays, the algorithm
investigates 1104 different decay topologies. The output variable $o_\textrm{tag}$ of the algorithm takes a value between 0 and 1, with larger values corresponding to a higher likelihood that a $B$ meson was correctly reconstructed. 

We select events with $\textrm{log}(o_\textrm{tag} ) > -3.5$. 
For each $B_{\rm tag}$, we impose a requirement on the difference between the measured center-of-mass (CM) energy and its nominal value of $|\Delta E | = | E_{B_{\textrm{tag}}}-E_{\textrm{CM}}| < 0.18 $\,GeV, and on the beam constrained mass of $M_{\textrm{bc}}=\sqrt{(E_\textrm{CM}/c^2)^2-(\vec{P}_{B_{\textrm{tag}}}/c)^2} >5.27$ GeV/$c^2$. Here, $E_{B_{\textrm{tag}}}$ and $\vec{P}_{B_{\textrm{tag}}}$ are the energy and momentum of the tagged $B$ candidate.

Differences in the tagging efficiency between data and MC have been observed~\cite{Sibidanov:2013rkk}. These depend on the tag-side reconstruction and the value of $o_\textrm{tag}$. We use a calibration derived in Ref.~\cite{Sibidanov:2013rkk}, which uses a control sample $B\rightarrow X_c l \nu$ decays on the signal side.
Based on this calibration, we assign an event-by-event weight based on the reconstructed $B_\textrm{tag}$ decay mode and value of $o_\textrm{tag}$ to equalize the efficiency of the tagging algorithm between data and MC.

\subsection{$B_\textrm{sig}$ reconstruction}
Having selected the $B_\textrm{tag}$ in this way, the signal side $B_\textrm{sig}$ is then reconstructed with the charged tracks and photons in the event that are not part of the $B_\textrm{tag}$ decay chain.
Charged tracks are identified using the Belle particle identification (PID)~\cite{Nakano:2002jw}. We accept electrons in the laboratory frame polar-angle range $17\,\degree < \theta_e < 150\,\degree$, and muons in the range $25\,\degree < \theta_\mu < 145\,\degree$, where the relevant subsystems of the Belle PID have acceptance for these particles.
To recover energy lost by bremsstrahlung of electrons, we add the 4-vector of the closest $\gamma$ found within 5\,$^\circ$ of an identified electron.
Charged tracks that cannot be unambiguously identified are treated as pions.
We reconstruct $\pi^0$ candidates from pairs of photons, each of which satisfies a minimum energy requirement of 50\,MeV, 75\,MeV or 100\,MeV in the barrel ($32\,\degree < \theta_\gamma<130\,\degree$), the forward endcap ($17\,\degree<\theta_\gamma<32\,\degree$) or the backward endcap ($130\,\degree<\theta_\gamma<150\,\degree$), respectively. We require the reconstructed mass to lie in the range $0.12$\,GeV/c$^2$ $< M_{\gamma \gamma}<0.15$\,GeV/c$^2$, which corresponds to about five times the measured resolution around the nominal mass. To reduce overlap in the $\pi^0$ candidate list, we sort them according to the most energetic daughter photon (and then, if needed, the second most energetic daughter) and remove any pion that shares photons with one that appears earlier in this list.  We reconstruct $K_S^0$ mesons from $\pi^+\pi^-$ pairs.  We require the two-pion invariant mass to lie in the range 0.482-0.514\,GeV/c$^2$ (about four times the experimental resolution around the nominal mass~\cite{Olive:2016xmw}).
Different selections are applied, depending on the momentum of the
$K_S^0$ candidate in the laboratory frame~\cite{Glattauer:2015teq}:  For low $(p<0.5$\,GeV/$c$), medium $(0.5 \leq p \leq  1.5$\,GeV/$c$),
and high momentum $(p > 1.5$\,GeV/c $)$ candidates, we require the impact
parameters of the pion daughters in the transverse plane (perpendicular to the beam) to be greater than 0.05\,cm, 0.03\,cm, and 0.02\,cm, respectively.  The angle in the transverse plane between the vector from the interaction point to the $K_S^0$ vertex and the $K_S^0$ flight direction is required to be less than 0.3\,rad, 0.1\,rad, and 0.03\,rad for low, medium, and high momentum candidates, respectively; the separation distance along the $z$ axis of the two pion trajectories at their closest approach must be below 0.8\,cm,  1.8\,cm,  and 2.4\,cm,  respectively.  Finally,  for medium (high) momentum $K_S^0$ candidates, we require the flight length in the transverse plane to be greater than 0.08\,cm (0.22\,cm).
Using the reconstructed pions and kaons, we reconstruct $D$ mesons in the channels $D^0\rightarrow K^-\pi^+$, $D^0\rightarrow K^-\pi^+\pi^0$, $D^0\rightarrow K^-\pi^+\pi^+\pi^-$, $D^0\rightarrow K_S^0\pi^+\pi^-$, $D^0\rightarrow K^-K^+$, $D^0\rightarrow K_S^0 \pi^0$, $D^+\rightarrow K_S^0 \pi^+$, $D^+\rightarrow K_S^0 \pi^+\pi^+\pi^-$, $D^+\rightarrow K^-\pi^+\pi^+$, and $D^+\rightarrow K^+K^-\pi^+$. Here and throughout this report, the charge-conjugated modes are implied.
We require a maximum difference of $3\sigma$ between the reconstructed mass and the nominal $D$ mass. This corresponds to 15~MeV for all modes except the $D^0\rightarrow K^-\pi^+\pi^0$ channel where the corresponding value is 25\,MeV.
Using the $D$ candidates, we reconstruct $D^*$ mesons in the channels $D^{*0}\rightarrow D^0 \pi^0$, $D^{*+}\rightarrow D^+ \pi^0$, and $D^{*+}\rightarrow D^0 \pi^+$.
The maximal difference allowed between the reconstructed mass and the nominal value is 3\,MeV, which again corresponds to $3\sigma$.
For both the $D$ and $D^*$ reconstruction, we perform a mass-vertex constrained fit and discard candidates for which this fit fails.
We require that no additional charged track be in the event other than the decay products of the $B_\textrm{tag}$, $D^{(*)}$, the lepton, and the signal's bachelor pion. Furthermore, we require the 
 lepton and bachelor pion to be positively identified. We require that the pion, lepton and $D^{(*)}$ meson form a overall charge neutral system with $B_\textrm{tag}$.
We also require $M_{D^{(*)}  \pi }$ to be less than 3\,GeV/c$^2$ and larger than 2.05\,GeV/c$^2$. 
There is the possibility of signal overlap, \textit{i.e.} the non-tag final state may be combined in different signal states. 
This overlap fraction is about 5\,\%. In such cases, we select at most one $B_\textrm{sig}$ candidate per event using two criteria. First, we prefer $D^{*}$ over $D$ in the final state since, otherwise, we would have an extra $\pi^0$ in the event, leading to additional missing energy.
Second, we select the $D^{(*)}$ whose reconstructed mass is closer to its nominal value.
The requirements described above for $M_{\textrm{bc}}$, $\Delta E$, $o_\textrm{tag}$, and $M_{D^{(*)}  \pi }$
are determined by maximizing the figure of merit $S/\sqrt{S+B}$ using MC simulation; here, $S$ and $B$ are the signal and background yields, respectively.

\subsection{Extraction of the branching fraction}

The branching fractions are determined by fitting the $M^2_\nu=\left((p_{e^+}+p_{e^-})-p_{B_\textrm{tag}}-p_{D^{(*)}}-p_\pi-p_l\right)^2/c^2$ spectrum.
Here, $(p_{e^+}+p_{e^-})$ is the sum of the four-momenta of the colliding beam particles and the other terms are the four-momenta of the indicated final-state particles.
We fit the spectrum with probability density function (PDF) templates derived from simulation to extract the yields; then we determine $\mathcal{B}$, using the ratios of the fitted yields to those in the original MC and the branching fractions used in MC.

The agreement of the simulations with data is checked by comparing the sidebands [$-1 $\,$(\textrm{GeV/}c^2)^2$ $ < M^2_\nu < -0.5$\,$(\textrm{GeV/}c^2)^2$ and $ 2 $\,$(\textrm{GeV/}c^2)^2$ $< M^2_\nu< 3.5$\,$(\textrm{GeV/}c^2)^2$] and the signal region for events that were discarded for failing to form a charge-neutral system. The reduced $\chi^2$, obtained by comparing the difference between data and MC, for these tests is $1.02$, showing that the agreement of data and MC is good.

For the channels $B^+\rightarrow D^{(*)-} \pi^+ \ell^+ \nu$  and $B^0\rightarrow \bar{D}^{(*)0} \pi^- \ell^+ \nu$, we consider the following components in the MC:
\begin{itemize}
\item $B\rightarrow  D\pi  \ell \nu$
\item $B \rightarrow D^* \pi \ell \nu$
\item $B\rightarrow  D^{(*)}\pi  \ell \nu$ where the charge of the $B$ meson is inconsistent with the charge of $B_\textrm{tag}$
\item  $B\rightarrow  D^{(*)} \pi \pi \ell \nu$ 
\item $B\rightarrow D^{(*)} \ell \nu$
\item other $B\bar{B}$
\item continuum contributions.
\end{itemize}

Since $B\rightarrow D^* \pi \ell\nu$ contributes also as feed-down 
 to $B\rightarrow D \pi \ell \nu$ with a known ratio, we fit simultaneously the $B \rightarrow D \pi \ell\nu$ and $B\rightarrow D^* \pi l\nu$ channels. Charged and neutral $B$ channels are fitted
 separately.
 
\begin{figure*}
\includegraphics[width=0.95\textwidth]{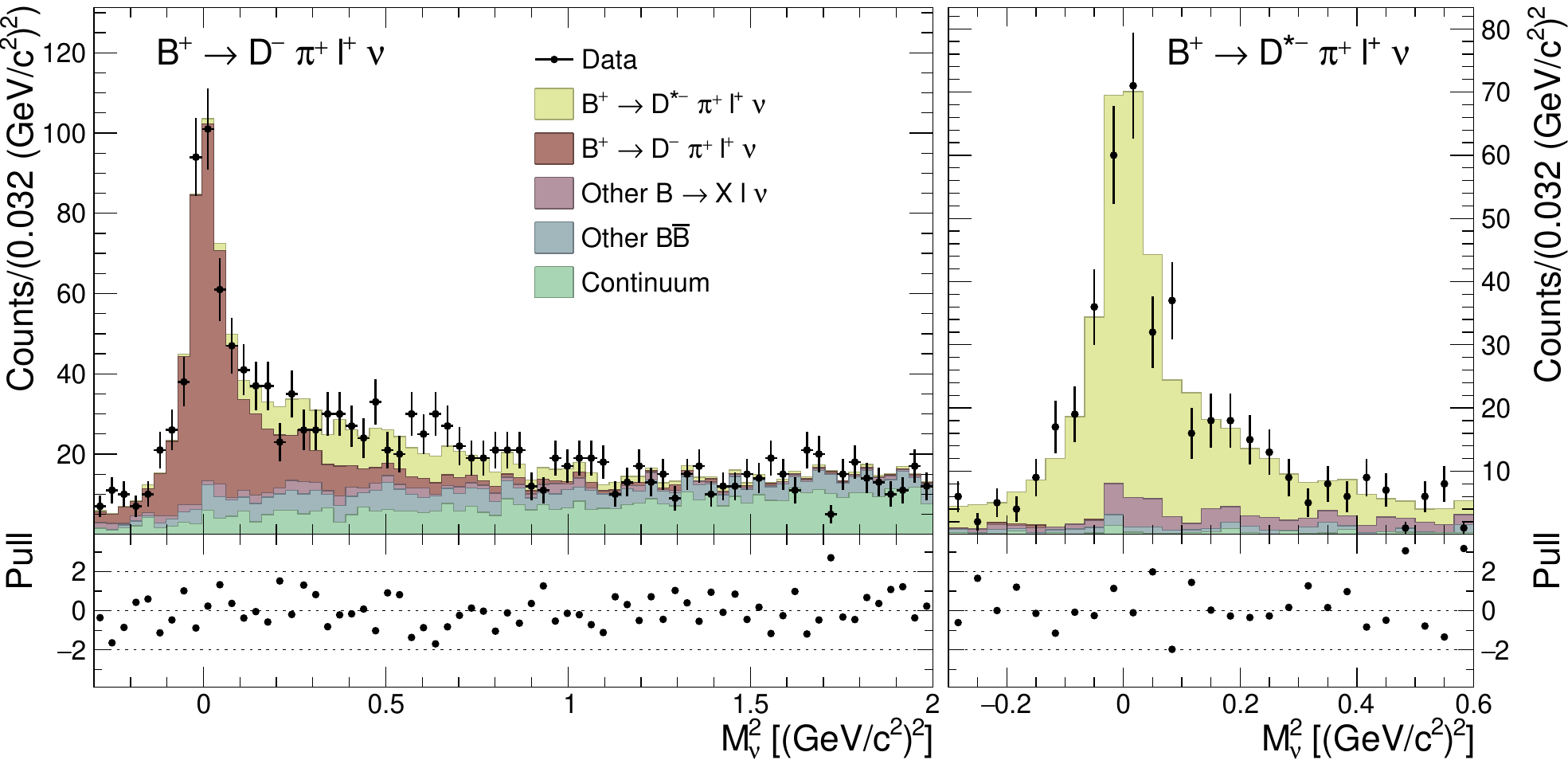}
\caption{(Color online) Binned extended maximum likelihood of the MC templates to the data for the combined fit to \bToD (left) and \bToDStar (right). The data is shown with error bars. The legend in the left panel indicates each component in the fit. The dots at the bottom of each panel show the pulls between the data and the fit. For better visibility, we doubled the bin width for this plot.\label{fig:resBToD}}
\end{figure*}

\begin{figure*}
\includegraphics[width=0.95\textwidth]{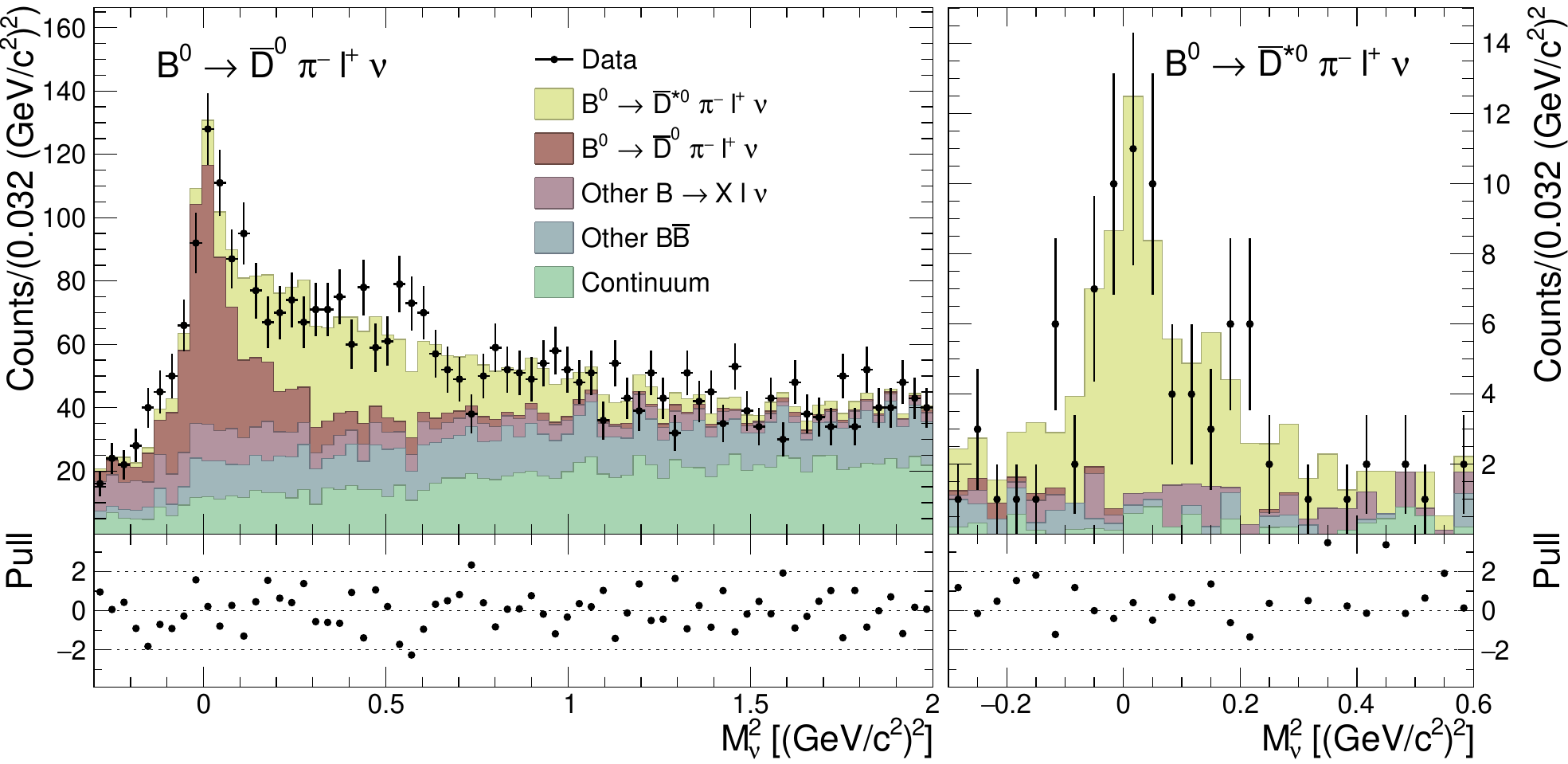}
\caption{(Color online) Binned extended maximum likelihood of the MC templates to the data for the combined fit to \bZToD (left) and \bZToDStar (right). The data is shown with error bars. The legend in the left panel indicates each component in the fit. The dots at the bottom of each panel show the pulls between the data and the fit. For better visibility, we doubled the bin width for this plot.\label{fig:resB0ToD}}
\end{figure*}

The simulation sample corresponds to five times the integrated luminosity of the data. With the given statistics, not all templates can be determined precisely enough for a stable fit. We therefore float only the $B\rightarrow D \pi \ell \nu$, the  $B \rightarrow D^* \pi \ell \nu$ and the continuum yields. The contribution from ``other $B\bar{B}$'' is not small; however, the shape is very similar to the continuum contribution and, given the agreement of the data and simulation in the sidebands, it is reasonable to fix this contribution to the MC prediction.
We use a binned extended maximum likelihood fit to extract the yields. The range of the fit is $-0.3$\,$(\textrm{GeV/c}^2)^2< M_\nu^2 < 2.0$\,$(\textrm{GeV/c}^2)^2$ with 140 bins for the \bToD and \bZToD channels. For the \bToDStar and \bZToDStar channels, we use a range of $-0.3$\,$(\textrm{GeV/c}^2)^2< M_\nu^2 < 0.6$\,$(\textrm{GeV/c}^2)^2$ with 54 bins. In the given $M_\nu^2$ ranges, we select
1566, 438, 3750, and 87 candidates for the $B^+ \rightarrow D^-\pi^+ \ell^+\nu$, $B^+ \rightarrow D^{*-}\pi^+ \ell^+\nu$, $B^0 \rightarrow \bar{D}^0\pi^- \ell^+\nu$, and \bZToDStar channels, respectively. Figure~\ref{fig:resBToD} shows the result of the fit to the combined \bToD and \bToDStar channels and Fig.~\ref{fig:resB0ToD} for the combined \bZToD and \bZToDStar channels. The $\chi^2/\textrm{Ndf}$ value for the $B^+$ and $B^0$ mode fits is 1.1 and 1.2, respectively. Ndf refers to the number of degrees of freedom in the fit. Since the counts for some entries in the fitted histograms are small, we use the equivalent quantity for Poisson statistics (see, \textit{e.g.}, Eq. (40.16) in Ref.~\cite{Olive:2016xmw}). Tables~\ref{tbl:fitResultsBToD} and~\ref{tbl:fitResultsBZToD} summarize the fit results.

\begin{table}
\ra{1.3}
\begin{tabular}{lS}
\toprule
 Source & {yield} \\ 
\colrule
$ B^+ \rightarrow  D^- \pi^+  \ell^+ \nu  $& 515+-31 \\

$ B^+ \rightarrow D^{*-} \pi^+ \ell^+ \nu$ & 571+-40\\
 Continuum & 444+-136 \\
 Other $B\bar{B}$ (fixed) & 360\\
 Other semi-leptonic $B$ decays (fixed) & 114\\
\botrule
\end{tabular}
\caption{Results for the combined fit \bToD and \bToDStar. \label{tbl:fitResultsBToD}}
\end{table}		

\begin{table}
\ra{1.3}
\begin{tabular}{lS}
\toprule
 Source & {yield} \\ 
\colrule
$B^0 \rightarrow \bar{D}^0 \pi^-  \ell^+ \nu $ & 537+-48\\
 $B^0  \rightarrow \bar{D}^{*0} \pi^- \ell^+ \nu $& 878 +- 72\\
 Continuum & 1164 +- 323\\
Other $B\bar{B} $ (fixed)& 856\\
Other semi-leptonic $B$ decays (fixed) & 401\\
\botrule
\end{tabular}
\caption{Results for the combined fit \bZToD and \bZToDStar. \label{tbl:fitResultsBZToD}}
\end{table}
We check that the fits are unbiased and give the expected uncertainty by fitting ensembles of simulated events generated by sampling from the fitting templates.
We plot the resulting residuals, fit them to a normal distribution, and check the mean and standard deviation.
Finally, we correct for the fact that our efficiency in $M_{D^{(*)}\pi}$ is not constant. Since in the simulation the shape of $M_{D^{(*)}\pi}$ is determined by the poorly-known widths and relative branching fractions of the $D^{**}$ mesons, it might be different in data. Therefore the non-constant efficiency may introduce an overall efficiency difference between data and simulation.  We use a quadratic function to fit the efficiency for each channel after determining that higher-order polynomials do not improve the fit quality significantly. Then we determine the shape of $M_{D^{(*)}\pi}$  in data by subtracting the background components determined from simulation using the $\mathcal{B}$ determined from our fit to $M_\nu^2$. Comparing the integrated efficiency in data and simulation for the signal, we determine overall-efficiency calibration factors of $1.008\pm 0.007$ for \bToDNS, $0.983\pm 0.006$ for \bZToDNS, $0.997\pm 0.002$ for \bToDStarNS, and $0.98\pm 0.01$ for \bZToDStarNS.

\subsection{Determination of systematics}
There are three main sources of systematic uncertainties for our measurement: uncertainties in the simulation of our detector and underlying physics process, the statistical uncertainties of our fitting templates and the uncertainty of the efficiency correction based on the $M_{D^{(*)}\pi}$ shapes in data and MC. 
For all three of these sources, our strategy to determine the systematic uncertainty is to use a MC approach that is based on running 1000 ensembles of simulated events, where the source of the systematic uncertainty is varied as described below for each source. We check that the refitted branching fraction in question follows a normal distribution and use the standard deviation of this distribution as our systematic uncertainty.

For the uncertainties of the simulation of the detector, we consider the uncertainty in the determination of the correction factors of the simulation of the PID discussed earlier as well as the uncertainty on the tracking efficiency. Similarly, for the underlying physical processes, we consider the uncertainty of the $D$ and $B$ meson branching fractions and the  $D^*$ and $D^{**}$ form factors. Furthermore, we consider the uncertainty of the calibration of the tagging algorithm, the uncertainty on the total number of $B\bar{B}$ pairs, and the uncertainty on the branching fractions of $\Upsilon(4S)$ to $B^+B^-$ and $B^0\bar{B^0}$. These sources of uncertainty of the simulation of the detector and underlying physical processes are described in more detail in Ref.~\cite{Glattauer:2015teq}. 
Since it is reasonable to assume that the sources of uncertainty follow a normal distribution, we draw for each ensemble of simulated events, source, and kinematic bin a new weight from a normal distribution with the corresponding width. This is then used to do an event-by-event weighting of the ensemble of simulated events. The advantage of this method is that correlations among the different sources for uncertainties as well as the dependence on the event kinematics are taken into account. By repeating this exercise while varying only one source at a time, we estimate the relative contributions of each source to the systematics. This decomposition is shown in Tables~\ref{tbl:mcUncertainties1} and~\ref{tbl:mcUncertainties2}. We omit the uncertainties due to the $K_S^0$ efficiencies and the $D^*$ form factors because these are consistent with zero relative to the tabulated uncertainties. 

\begin{table}
\ra{1.3}
\begin{tabular}{lSS}
\toprule
 &{\bToD}& {\bZToD} \\
\colrule
Charged PID & 4.8  & 6.9 \\
$\pi^0$ PID &  1.2 & 6.0 \\
Tracking efficiency &  2.6 & 3.6    \\
$D^{**}$ form factors & 0.3 & 0.2 \\
$D$ meson BRs &  1.7 & 1.6 \\
$B$ meson BRs & 0.0 & 0.1 \\
Number of $B\bar{B}$  & 1.4 & 1.4 \\
Tag efficiency & 4.6 & 3.2 \\
$\Upsilon(4S)$ BR & 1.2 &1.2\\
\colrule
{\bf Combined (see text)}  & {\bf 8.3} & {\bf 9.7}  \\
\botrule

\end{tabular}
\caption{Sources of uncertainty in the MC simulations considered for systematic uncertainties for the channels \bToD and \bZToD.
The table lists the relative uncertainties in the branching fractions in percent for each channel for the combined fits. The last row gives the combined variation of all sources. 
\label{tbl:mcUncertainties1}}
\end{table}
\begin{table}
\ra{1.3}
\begin{tabular}{lSS}
\toprule
& {\bToDStar} & {\bZToDStar}\\
\colrule
Charged PID  & 2.1  & 6.5\\
$\pi^0$ PID  & 2.0 & 5.2\\
Tracking efficiency & 2.9 &3.2   \\
$D^{**}$ form factors & 	0.2 & 0.1\\
$D$ meson BRs  & 1.8 & 1.1\\
$B$ meson BRs& 0.0 & 0.1\\
Number of $B\bar{B}$  & 1.4 & 1.4\\
Tag efficiency & 4.2 & 2.8\\
$\Upsilon(4S)$ BR & 1.2 &1.2\\

\colrule
{\bf Combined (see text)}   &{\bf 5.8} &{\bf 7.2} \\
\botrule

\end{tabular}

\caption{Sources of uncertainty in the MC simulations considered for systematic uncertainties for the channels \bToDStar and\bZToDStar.
The table lists the relative uncertainties in the branching fractions in percent for each channel for the combined fits. The last row gives the combined variation of all sources.
\label{tbl:mcUncertainties2}}
\end{table}

From Tables \ref{tbl:mcUncertainties1} and \ref{tbl:mcUncertainties2}, the combined systematic uncertainties on the branching fraction by varying all sources simultaneously are $8.3$\,\% for \bToDNS, $9.7$\,\% for \bZToDNS, $5.8$\,\% for \bToDStarNS, and $7.2$\,\% for \bZToDStarNS. 

We estimate the systematic uncertainties propagated from the statistical uncertainty of the fitting templates to be 1.9\%, 2.6\%, 3.2\%, and 3.5\% for the \bToD, \bToDStar, \bZToD and \bZToDStar \ channels, respectively. These values are estimated using 1000 ensembles of simulated events for which we vary the templates using Poisson statistics.
Finally, the uncertainty on the detector-efficiency dependence on $M_{D^{(*)}\pi}$ is estimated by varying the $M_{D^{(*)}\pi}$ spectrum for each channel within Poisson statistics and adding the difference of the average efficiency between the $\pm 68$\,\% boundaries of the fit to the efficiency versus $M_{D^{(*)}\pi}$. The resulting uncertainty propagated to the branching fraction of interest is below 1\.\% for each channel.
The final systematic uncertainties on the branching fraction from all sources discussed above correspond to 8.6\,\%  for \bToDNS, 10.3\,\% for \bZToDNS, 6.4\,\% for  \bToDStarNS, and 8.0\,\% for \bZToDStarNS.

\section{Results and conclusion}
Using the combined fits, including the correction and systematics from the $M_{D^{(*)}\pi}$ efficiency, simulation uncertainties and statistical uncertainty of the templates, we obtain the following values for the branching fractions: 
\begin{itemize}
\item $\mathcal{B}($\bToDNS$)$ \\= [4.55 $\pm$ 0.27 (stat.) $\pm$ 0.39 (syst.)] $ \times 10^{-3}$,
\item $\mathcal{B}($\bZToDNS$)$ \\= [4.05 $\pm$ 0.36 (stat.) $\pm$   0.41 (syst.)]$ \times 10^{-3}$,
\item $\mathcal{B}($\bToDStarNS$)$ \\= [6.03 $\pm$ 0.43 (stat.) $\pm$  0.38 (syst.)]$ \times 10^{-3}$,
 \item $\mathcal{B}($\bZToDStarNS$)$ \\= [6.46 $\pm$ 0.53 (stat.) $\pm$  0.52 (syst.)]$ \times 10^{-3}$.
 \end{itemize}
 
  These are within one standard deviation of the current world-average values~\cite{Olive:2016xmw} with the exception of \bZToDStar, which deviates by $1.7\sigma$. These supersede the previous Belle result~\cite{Liventsev:2007rb}. 
  The total uncertainties on our measurement are slightly better than the current world-average for the channels \bZToD and \bZToDStar, whereas they are the same for the channels \bToD and \bToDStarNS. A potential extension to this work would be to confirm the recent observation of $ B\rightarrow D^{(*)}\pi\pi \ell \nu$ by BaBar~\cite{Lees:2015eya} as well as to analyze the $M_{D^{(*)}\pi}$ distribution to extract the branching fractions and widths of the different $D^{**}$ mesons for which there are still some discrepancies between the Belle~\cite{Liventsev:2007rb} and BaBar~\cite{Aubert:2008ea} measurements.
\begin{acknowledgments} 
We thank the KEKB group for the excellent operation of the
accelerator; the KEK cryogenics group for the efficient
operation of the solenoid; and the KEK computer group,
the National Institute of Informatics, and the 
Pacific Northwest National Laboratory (PNNL) Environmental Molecular Sciences Laboratory (EMSL) computing group for valuable computing
and Science Information NETwork 5 (SINET5) network support.  We acknowledge support from
the Ministry of Education, Culture, Sports, Science, and
Technology (MEXT) of Japan, the Japan Society for the 
Promotion of Science (JSPS), and the Tau-Lepton Physics 
Research Center of Nagoya University; 
the Australian Research Council;
Austrian Science Fund under Grant No.~P 26794-N20;
the National Natural Science Foundation of China under Contracts
No.~11435013,  
No.~11475187,  
No.~11521505,  
No.~11575017,  
No.~11675166,  
No.~11705209;  
Key Research Program of Frontier Sciences, Chinese Academy of Sciences (CAS), Grant No.~QYZDJ-SSW-SLH011; 
the  CAS Center for Excellence in Particle Physics (CCEPP); 
Fudan University Grant No.~JIH5913023, No.~IDH5913011/003, 
No.~JIH5913024, No.~IDH5913011/002;                        
the Ministry of Education, Youth and Sports of the Czech
Republic under Contract No.~LTT17020;
the Carl Zeiss Foundation, the Deutsche Forschungsgemeinschaft, the
Excellence Cluster Universe, and the VolkswagenStiftung;
the Department of Science and Technology of India; 
the Istituto Nazionale di Fisica Nucleare of Italy; 
National Research Foundation (NRF) of Korea Grants No.~2014R1A2A2A01005286, No.2015R1A2A2A01003280,
No.~2015H1A2A1033649, No.~2016R1D1A1B01010135, No.~2016K1A3A7A09005 603, No.~2016R1D1A1B02012900; Radiation Science Research Institute, Foreign Large-size Research Facility Application Supporting project and the Global Science Experimental Data Hub Center of the Korea Institute of Science and Technology Information;
the Polish Ministry of Science and Higher Education and 
the National Science Center;
the Ministry of Education and Science of the Russian Federation and
the Russian Foundation for Basic Research;
the Slovenian Research Agency;
Ikerbasque, Basque Foundation for Science, Basque Government (No.~IT956-16) and
Ministry of Economy and Competitiveness (MINECO) (Juan de la Cierva), Spain;
the Swiss National Science Foundation; 
the Ministry of Education and the Ministry of Science and Technology of Taiwan;
and the United States Department of Energy and the National Science Foundation.
\end{acknowledgments}

\bibliography{Blnu}
\end{document}